\documentstyle[pra,aps,multicol,eqsecnum,epsfig]{revtex}
\newcommand\be{\begin{eqnarray}}
\newcommand\ee{\end{eqnarray}}
\newcommand\ba{\begin{array}}
\newcommand\ea{\end{array}}
\def\r{\rangle}
\def\l{\langle}
\def\T{{\rm Tr}}

\def\cH{{\cal H}}
\def\cS{{\cal S}}

\def\cI{{\cal I}}

\def\cF{{\cal F}}

\def\cE{{\cal E}}

\def\cD{{\cal D}}

\def\cG{{\cal G}}

\def\openone{{\it I}}

\begin{document}
\title{Reconstruction of superoperators from incomplete measurements
\thanks{We dedicate this paper to Asher Peres on the occasion of his 70th birthday}
}
\author{M\'ario Ziman$^{1}$, Martin Plesch$^{1}$, and Vladim\'\i r Bu\v zek$^{1,2}$}
\address{$^{1}$Research Center for Quantum Information, Institute of Physics, Slovak Academy of Sciences,
D\'{u}bravsk\'{a} cesta 9, 845 11 Bratislava, Slovakia\\
$^{2}$Faculty of Informatics, Masaryk University, Botanick\'a 68a,
602 00 Brno, Czech Republic}
\date{19 March 2004}
\maketitle

\begin{abstract}
We  present strategies how to reconstruct (estimate)
properties of a quantum channel described by the map $\cE$ based on incomplete measurements.
In a particular case
of a qubit channel a complete reconstruction of the map $\cE$ can be performed
via complete tomography of four output states $\cE[\varrho_j]$ that originate from
a set of four linearly independent ``test'' states $\varrho_j$ ($j=1,2,3,4$) at the input of the channel.
We study the situation when less than four linearly independent states are transmitted
via the channel and measured at the output. We  present strategies how to reconstruct
the channel when just one, two or three states are transmitted via the channel. In particular,
we show that if just one state is transmitted via the channel then the best
reconstruction can be achieved when this state is a total mixture described by the density
operator $\varrho = \openone /2$. To improve the reconstruction procedure one has to send via the channel
more states. The best strategy is to complement the total mixture with
pure states that are mutually orthogonal in the sense
of the Bloch-sphere representation.  We show that unitary transformations (channels) can be
uniquely reconstructed (determined) based on the information of how three properly chosen input states
are transformed under the action of the channel.
\end{abstract}
\bigskip\bigskip
\hfill
\begin{minipage}[r]{5.5cm}
{\it ... "true physical situation" ... is unknown, and always remains
so, because of incomplete information.
} \rightline{E. T. Jaynes
\cite{Jaynes1962}}
\end{minipage}
\bigskip
\bigskip

\begin{multicols}{2}
\section{Introduction}
Our observation of external world is always incomplete and the main task of a physicist
is to find the best possible description  of our state of knowledge about the physical situation.
A typical example of this  is a description of the action of an unknown quantum channel $\cE$.
Specifically,
it is well known that a complete  correlation measurement between a properly chosen
set of linearly independent input states $\rho_j$ and corresponding outputs $\cE [\rho_j]$ reveal
the complete information about the character of the map $\cE$ (see e.g.~Ref.\cite{Poyatos,Ariano,Ariano2}).
The question is, how to describe the map $\cE$ if only results of incomplete correlation measurements
are known? In this paper we will address some aspects of this problem.

Let us formulate our problem in more detail:
In quantum theory (under reasonable conditions) the evolution
map is associated with a trace-preserving completely positive
linear map $\cE$ \cite{Peres1995,Nielsen} acting on a set of quantum
states $\cS(\cH)$. The set of states is a subset (not subspace)
of the real linear space of all hermitian operators with dimension $d^2$,
where $d={\rm dim}\cH$. To define a linear map $\cE$ we have to specify
its action on an operator basis, i.e. we have to specify
$d^2\times d^2$ parameters \cite{Poyatos}. In addition, the condition that
the map $\cE$ is trace-preserving reduces this number to $d^4-d^2$.
From here it follows that in order  to determine a transformation $\cE$ experimentally we have
to choose $d^2$ {\em linearly independent} states $\rho_j$. These states are sent via the channel $\cE$.
At the output of the channel we
perform a complete state reconstruction of the  states $\cE [\rho_j]$. Given the fact that each of the $d^2$ states
is determined by $d^2-1$ real numbers we clearly see that the map $\cE$ can be completely determined via this
so-called {\it quantum process tomography}.

However, also another scenario is possible \cite{Ariano,Ariano2,jezek,raginski}:
One can use a {\em single} maximally
entangled state to probe the action of the channel. In particular, let us
consider a maximally entangled state
$|\psi_+\r=\frac{1}{\sqrt{d}}\sum_j |j\r\otimes|j\r$ from the Hilbert space
$\cH\otimes\cH$. One of the subsystem is transformed by the channel,
while the other (reference) system evolves freely without any disturbance. In this case the final state of the
composite system reads
\be
\Omega = \cE\otimes\cI[P_+]=\frac{1}{d}\sum_{jk}\cE[e_{jk}]\otimes e_{jk}\, ,
\ee
where $P_+=|\psi_+\r\l\psi_+|=\frac{1}{d}\sum_{jk} e_{jk}\otimes e_{jk}$ and operators
$e_{jk}=|j\r\l k|$ form an operator basis of the operator space.
Via the reconstruction of  the output state $\Omega$ we obtain the complete information
about the map $\cE$. This is due to the fact that the input-output correlation between
states $P_+$ and $\Omega$ provides us with the knowledge how the
operator basis is transformed.

In what follows we will
 consider the first reconstruction scenario (the one that does not involve entangled states) but
we study the situation when the number of linearly independent input states is less than $d^2$. In this situation
the perfect process tomography cannot in general be performed.
To simplify our analysis  we
will assume that the preparation of test states $\rho_j$ as well as the determination of the output state
can be performed perfectly (e.g., via quantum-state tomography).

The present article is organized as follows: In Sec.~II we briefly review  basic formalism
of completely positive maps that act on a qubit. For the completeness of our discussion we will also
show how the CP map can be completely reconstructed.
In Sec.~III we will present our strategy
how to perform reconstruction of quantum processes from incomplete measurements
and we will analyze number of examples. We conclude our paper
by several comments that can be found in Sec.~IV.

\section{Properties of qubit quantum operations}

In what follows we will focus our attention on a reconstruction
of single-qubit
transformations - i.e., CP maps on  {\it qubit channels}. The state space ${\cal S}({\cal H})$
is a convex subset (not subspace) of a four-dimensional linear space of hermitian
operators. This means that any state can be expressed as a four-dimensional real vector.
Due to linearity of quantum mechanics, the state transformation $\cE$ can be
represented by a 4x4 matrix. If we use the basis of $\sigma$
matrices
($\sigma_0=\openone,\sigma_1=\sigma_x,\sigma_2=\sigma_y,\sigma_3=\sigma_z$),
then an arbitrary state of a qubit can be expressed as
\begin{eqnarray}
\varrho=\frac{1}{2}(\openone+\vec{r}\cdot\vec{\sigma})=\frac{1}{2}\left(
\begin{array}{c}
1 \\
x \\
y \\
z
\end{array}
\right)\leftrightarrow\vec{r}=(x,y,z)\, ,
\end{eqnarray}
with $x^2+y^2+z^2\le 1$. From here it follows that the state space of a qubit  can be
represented (visualized) as a
three-dimensional (Bloch) sphere. The center  $\vec{r}=(0,0,0)$
of the Bloch sphere corresponds to an equally weighted
superposition of all qubits states, i.e. it represent a {\it total mixture}.
Let us note that the Bloch sphere (state space)
form a subset of a three-dimensional space of
hermitian matrices with unit trace. In what follows we will often omit
the factor $1/2$ in the vector representation.

In this basis any evolution ${\cal E}$
takes the affine form
\begin{eqnarray}  \label{super}
{\cal E}=\left(
\begin{array}{cc}
1 & 0 \\
\vec{t} & E
\end{array}
\right)\, ,
\end{eqnarray}
where $E$ is a real 3x3 matrix. In particular,
\begin{eqnarray}
\nonumber
\vec{r}\to\vec{r}^{\,\prime}=E\vec{r}+\vec{t} \, .
\end{eqnarray}
The vector $\vec{t}$ corresponds to a translation of the center of the Bloch
sphere. If $\vec{t}=\vec{0}$,
then the total mixture is preserved and the map ${\cal E}$ is called to be {\it %
unital}, or equivalently {\it bistochastic} \cite{Nielsen}.
For a unitary transformation the
matrix $E$ corresponds to a
three-dimensional orthogonal rotation, i.e. $E^{-1}=E^T$.
The Bloch sphere picture enables us to illustrate  quantum operations as
deformations of the unit sphere \cite{Nielsen,Ruskai,King}.

The  matrix form (\ref{super}) of $\cE$ guaranties the condition of  the trace preservation (first row).
As we have already mentioned the matrix $\cE$ is specified by $2^4-2^2=12$
real parameters, but because of the constraint of the complete positivity  the range
of these parameters is restricted. The transformation $\cE$ is completely positive if and
only if it can be written in the Kraus form \cite{Nielsen}
\be
\cE[\varrho]=\sum_k A_k\varrho A_k^\dagger\, ,\ \ {\rm with}\ \ \sum_k A_k^\dagger A_k=\openone \, ,
\ee
or equivalently \cite{Choi}
if the operator $\Omega=\cE\otimes\cI[P_+]$ is positive and if ${\rm Tr}_1 \Omega = \openone/2$, where
${\rm Tr}_1$ corresponds to a partial trace over the first system.
For completely positive maps the following relations hold \cite{Nielsen}:
\begin{enumerate}
\item{\it Contractivity of distance}
\be
D(\varrho,\xi) \ge D(\cE[\varrho],\cE[\xi])\, ,
\ee
where $D(A,B)=\T|A-B|$. For qubits this condition reads
\be
D(\varrho,\xi)=|\vec{r}-\vec{s}| \, ,
\ee
where $\varrho\leftrightarrow\vec{r}$ and $\xi\leftrightarrow\vec{s}$.
\item{\it Fidelity monotone}
\be
F(\varrho,\xi)\le F(\cE[\varrho],\cE[\xi]) \, ,
\ee
where $F(A,B)=\T\sqrt{\sqrt{A}B\sqrt{B}}$ is the fidelity.
\end{enumerate}

For our purposes it is useful to present explicitly the distance
and the fidelity between any state $\varrho$ and a {\it total mixture}
$\frac{1}{2}\openone\leftrightarrow\vec{m}=(0,0,0)$, i.e. the center
of the Bloch sphere. This distance equals
\be
\nonumber
D(\varrho,\frac{1}{2}\openone)&=&|\vec{r}|\, , \\
F(\varrho,\frac{1}{2}\openone)&=&\frac{1}{\sqrt{2}}\T\sqrt{\varrho}
= \frac{1}{2}(\sqrt{1+|\vec{r}|}+\sqrt{1-|\vec{r}|}) \, .
\nonumber
\ee
Both of these equations are monotone in the state $\varrho$.
The fidelity decreases as the $|\vec{r}|$ increases and the trace distance increases together
with $|\vec{r}|$. For non-unital maps ($\cE[\frac{1}{2}\openone]\ne\frac{1}{2}\openone$)
there are states for which  $D(\varrho,\frac{1}{2}\openone)<D(\cE[\varrho],\frac{1}{2}\openone)$,
or $F(\varrho,\frac{1}{2}\openone)>F(\cE[\varrho],\frac{1}{2}\openone)$.
Therefore if we find such states then we can conclude
the map $\cE$ must be non-unital.

Any state transformation $\cE$ can be (uniquelly) expressed
as a composition of  maps \cite{Ruskai}
\begin{eqnarray}
{\cal E}[\varrho]={\cal U}{\cal D}{\cal V}[\varrho]=U{\cal D}[V\varrho
V^\dagger]U^\dagger \, ,
\end{eqnarray}
where $U,V$ are unitary operations and ${\cal D}$ is a map of the form
(\ref{super})
with a diagonal matrix $D={\rm diag}\{\lambda_1,\lambda_2,\lambda_3\}$.
The $\lambda$'s are singular values of the matrix $E$
associated with the original map $\cE$, i.e. the squares of the eigenvalues of the matrix $EE^\dagger$.
We should note that also the vector
$\vec{t}$ is changed according to the rule $\vec{t}\to R^T_U\vec{t}$ where $R_U$ is
a rotation corresponding to the unitary transformation $U$. The existence of such
representation follows from the {\it polar decomposition theorem}. According to this theorem
matrices $E$ can be expressed in the form $E=R_U D R_V$ where $R_U,R_V$ are orthogonal rotations.
This means that
the conditions due to the complete positivity of the map $\cE$ are reduced
to the requirement of the complete positivity  of the map $\cD$.
Unfortunately, even for a qubit the conditions of complete positivity of $\cD$
is quite complex. The necessary conditions are \cite{Fujiwara,Ruskai}
\begin{eqnarray}
(\lambda_1\pm\lambda_2)^2\le (1\pm \lambda_3)^2-t_3^2 \, .
\end{eqnarray}
If $t_1=t_2=0$, then these conditions are also sufficient ones.
Consequently for unital maps
($\vec{t}=\vec{0}$)
we have the necessary and sufficient conditions in a simple form
\be
\label{unitalsuf}
(\lambda_1\pm\lambda_2)^2\le (1\pm \lambda_3)^2 \, .
\ee

The set of all completely positive linear trace-preserving maps
(quantum channels)
is convex. That is, we can define ``pure'' quantum channels as
{\em extremal} points of this set. Obviously, unitary transformations
are extremal. However, there exist also non-unitary
extremal quantum operations \cite{Ruskai}.
For instance, a contraction to single pure state, i.e.
$\varrho\to|\psi\r\l\psi|$ for all $\varrho$, is extremal.

In what follows we will often  use the following basis of the state space
\begin{eqnarray}
\nonumber
\varrho_m=\left(
\begin{array}{c}
1 \\
0 \\
0 \\
0
\end{array}
\right)
;\,
\varrho_x=\left(
\begin{array}{c}
1 \\
1 \\
0 \\
0
\end{array}
\right)
;\,
\varrho_y=\left(
\begin{array}{c}
1 \\
0 \\
1 \\
0
\end{array}
\right)
;\,
\varrho_z=\left(
\begin{array}{c}
1 \\
0 \\
0 \\
1
\end{array}
\right)\,
.
\end{eqnarray}
These states are eigenstates of the operators $S_x,S_y,S_z$
 associated with the value $+1$. The state $m$ describes the
total mixture. The operators $S_j$ are the generalized $\sigma$-matrices
with respect to a qubit state $|\psi\r$. They are defined by the following
relations
\be
\nonumber
S_x &=& |\psi\r\l\psi_\perp|+|\psi_\perp\r\l\psi|\, ,\\
S_y &=& -i(|\psi\r\l\psi_\perp|-|\psi_\perp\r\l\psi|)\, ,\\
S_z &=& |\psi\r\l\psi|-|\psi_\perp\r\l\psi_\perp|\, .
\nonumber
\ee
In what follows we will use the operator basis $\{S_j\}$
which is obtained by a unitary transformation of the original
$\sigma$-basis. This basis can be  illustrated as a set
of orthogonal axes in the Bloch-sphere picture. The above
basis is  determined by
the choice of cartesian coordinates in a three-dimensional space.

With the help of
the four states $\varrho_m,\varrho_x,\varrho_y,\varrho_z$
we can perform the complete reconstruction of the map $\cE$.
The reconstruction of $\cE$ requires the knowledge of the transformation of these states, i.e.
 $\cE[\varrho_m],\cE[\varrho_x],\cE[\varrho_y],\cE[\varrho_z]$.
These output states determines columns of the matrix $\cE$.
In particular, using the correspondence
$\cE[\varrho_x]\leftrightarrow\vec{x}^\prime$ (similarly for other basis
operators) we can write $\vec{t}=\vec{m}^\prime$ and the matrix $E$ reads
$E=(\vec{x}^\prime-\vec{t},\vec{y}^\prime-\vec{t},\vec{z}^\prime-\vec{t})$, so
the transformation $\cE$ can be expressed as
\be
\cE=\left(
\ba{cccc}
1 & 0 & 0 & 0 \\
\vec{m}^\prime &
\vec{x}^\prime-\vec{m}^\prime &
\vec{y}^\prime-\vec{m}^\prime &
\vec{z}^\prime-\vec{m}^\prime
\ea
\right)\, .
\ee
In general we can use any four linearly independent states for reconstruction purposes. Nevertheless,
it should be noted that it is always possible to transform  the reconstruction procedure  into the basis
discussed above.

As we have already mentioned, instead of using four linearly independent states of a qubit, one can use a single
entangled state$P_+$ of two qubits to perform a perfect reconstruction of the channel $\cE$.
Specifically, from the output state $\Omega={\cal E}\otimes{\cal I}[P_+]$ we can determine the map $\cE$.

In the $\sigma$ basis the state $P_+$ reads
\begin{eqnarray}
P_+=\frac{1}{4}({\it 1}\otimes{\it 1}+\sigma_x\otimes\sigma_x-
\sigma_y\otimes\sigma_y+\sigma_z\otimes\sigma_z)\, .
\end{eqnarray}
The output state $\varrho$ can be uniquely written in the form
\begin{eqnarray}
\nonumber
\Omega=\frac{1}{4}(\Omega_0\otimes{\it 1}+
\Omega_x\otimes\sigma_x+\Omega_y\otimes\sigma_y+\Omega_z\otimes\sigma_z)\, ,
\end{eqnarray}
where
\begin{eqnarray}
\nonumber
\Omega_0 &=&{\cal E}[{\it 1}] = \frac{1}{2}({\it 1}+\vec{t}\cdot\vec{\sigma}
)\, ; \\
\nonumber
\Omega_x &=&{\cal E}[\sigma_x] = \frac{1}{2}\vec{e}_x\cdot\vec{\sigma}\, ; \\
\nonumber
\Omega_y &=&-{\cal E}[\sigma_y]= - \frac{1}{2}\vec{e}_y\cdot\vec{\sigma}\, ; \\
\nonumber
\Omega_z &=&{\cal E}[\sigma_z] =\frac{1}{2}\vec{e}_z\cdot\vec{\sigma}\, .
\end{eqnarray}
The reconstruction of the state $\Omega$ gives us the operators $\Omega_k$
and the map ${\cal E}$ is completely determined by the above equations.
In particular, the vectors $\vec{
e}_k$ (with $\vec{e}_0=\vec{t}$) correspond to columns of the matrix
${\cal E}$, i.e.
\begin{eqnarray}
{\cal E}=\left(
\begin{array}{cccc}
1 & 0 & 0 & 0 \\
\vec{t} & \vec{e}_x & \vec{e}_y & \vec{e}_z
\end{array}
\right)\, .
\end{eqnarray}

Before we conclude this section we note that
in general, we do not need maximally entangled states to perform the
complete reconstruction of the map $\cE$ (for details see Ref.~\cite{Ariano,Ariano2}. In fact,
any state that can be written in the form
\begin{eqnarray}
\varrho= \frac{1}{4}\left( {\it 1}\otimes{\it 1}+x\sigma_x\otimes\sigma_x+
y\sigma_y\otimes\sigma_y+z\sigma_z\otimes\sigma_z \right)
\end{eqnarray}
can be used for the reconstruction, because the output state reads
\begin{eqnarray}
\nonumber
\Omega= \frac{1}{4}\left( {\cal E}[{\it 1}]\otimes{\it 1}+x{\cal E}%
[\sigma_x]\otimes\sigma_x+ y{\cal E}[\sigma_y]\otimes\sigma_y+z{\cal E}%
[\sigma_z]\otimes\sigma_z \right)
\end{eqnarray}
and the one-to-one correspondence between ${\cal E}$ and $\Omega$ is
obvious. In particular, every pure state can be written in the
Schmidt form $|\psi\rangle=a|00\rangle+b|11\rangle$,
where $\{|0\rangle,|1\rangle\}$ are suitable chosen bases of subsystems. If
we define the operators $S_x=|0\rangle\langle 1|+|1\rangle\langle 0|$, $%
S_y=-i|0\rangle\langle 1|+i|1\rangle\langle 0|$, $S_z=|0\rangle\langle
0|-|1\rangle\langle 1|$, then
\begin{eqnarray}
\nonumber
\Omega=\frac{1}{4} \left( {\it 1}\otimes{\it 1}+2ab S_x\otimes S_x-2ab
S_y\otimes S_y +S_z\otimes S_z \right)\, .
\end{eqnarray}
Consequently, whenever the pure state is correlated (entangled), it can be
used to perform the complete reconstruction of the CP map.

\section{Strategies for incomplete reconstruction}
Any reliable reconstruction scheme should be  as conservative as possible,
i.e. we cannot prefer any type of transformations that is not warranted by the data
\cite{Jaynes1962}.

Having this principle in mind
let us try to guess the expression for the map (i.e. the transformation) when no measurement
is performed. In the absence of measurement data arbitrary test state at the output
of the channel can be estimated as a
total mixture described by the density operator $\openone/2$ (for more details see \cite{Jaynes1962})
Consequently,
the estimated map describing the action of the channel describes a contraction of the whole Bloch sphere into
a single point - the total mixture, i.e. $\cE_0:\varrho\mapsto\frac{1}{2}\openone$.

The basic idea of our strategy is simple. If we have no {\em prior} information
about the action of the channel and  the data do not contain
any information about the transformation of a state $\varrho$, then we
assume that any input state is transformed by the channel into the total mixture, i.e.
$\cE[\varrho]=\frac{1}{2}\openone$. However, we have to clarify one point:
If we know {\em a priori} that the total mixture is affected by the channel $\cE$, i.e.
$\cE[\frac{1}{2}\openone]\ne \frac{1}{2}\openone$, then we
cannot assume that the unknown input  $\varrho$ is mapped into the total mixture.
We have to  assume that it is mapped as
 $\cE[\varrho]=\cE[\frac{1}{2}\openone]$. Of course, our guess
must satisfy constraints imposed by the complete positivity. Let us now present our
strategy:
\begin{enumerate}
\item{\it Step 1.} Check whether the available data contain
any information about the transformation of the total mixture, i.e.
whether we know $\cE[\frac{1}{2}\openone]$.
\item{\it Step 2.} If we have no information about a ``shift''
of the total
mixture, then we must check whether the map $\cE$ is unital
providing that all unknown states are mapped into $\frac{1}{2}\openone$.
In other words, check the complete positivity of the map $\cE$ when
all free parameters are equal to zero. If yes, our guess is completed.
\item{\it Step 3.} Choose the free parameters such that $\cE$
is completely positive, $D(\cE[\frac{1}{2}\openone],\frac{1}{2}\openone)$
is minimal and $D(\cE[\varrho_j],\cE_{\rm min}[\frac{1}{2}\openone])$ is also
minimal for all undetermined states $\varrho$.
\end{enumerate}

In what follows we will analyze reconstruction of
qubit channels. We will study three relevant cases: When we know how a single state
is transformed, when we know how two (linearly independent) states are transformed
and finally, when we known how three states are transformed.

\subsection{No input state}
Our guess is very simple in this case. Our strategy implies that all inputs are
mapped onto the total mixture. So the whole Bloch sphere is contracted
into its center and the map reads
\be
\cE_0=\left(
\ba{cccc}
1 & 0 & 0 & 0 \\
0 & 0 & 0 & 0 \\
0 & 0 & 0 & 0 \\
0 & 0 & 0 & 0
\ea
\right)\, .
\ee
\subsection{Single input state}
In this case we have to consider three different cases. Either the input state
is the total mixture, or it is a pure state, or an arbitrary mixed state.

Let us assume that our information is the following:
$\frac{1}{2}\openone\to\frac{1}{2}(\openone+\vec{r}^\prime\cdot\vec{\sigma})$, that is
we know that the input state that has been prepared in a total mixture has been transformed into
the state $\frac{1}{2}(\openone+\vec{r}^\prime\cdot\vec{\sigma})$. In this case
our guess of the transformation of other basis operators
is given by
\be
\cE[\varrho_x]=\cE[\varrho_y]=\cE[\varrho_z]=
\cE[\frac{1}{2}\openone]=
\frac{1}{2}(\openone+\vec{r}^{\,\prime}\cdot\vec{\sigma})\, .
\nonumber
\ee
Consequently, the reconstructed quantum channel $\cE$ is the contraction
to the point $\cE[\frac{1}{2}\openone]$
which is obviously a completely positive map.

Consider the case, when we know how a pure state $|\psi\r$
is transformed, i.e.
\be
\nonumber
|\psi\r\l\psi|=\frac{1}{2}(\openone+S_z)\to
\frac{1}{2}(\openone+\vec{r}^{\,\prime}\cdot\vec{S})=
\frac{1}{2}(\openone+ r^\prime S^\prime_z)\, ,
\ee
where $r^\prime=|\vec{r}^{\,\prime}|$.
We can assume that the map $\cE$ is unital, because
$D(\cE[P_\psi],\frac{1}{2}\openone)=r^\prime\le 1
= D(\frac{1}{2}\openone,P_\psi)$.
The guess is that $\cE[\varrho_x]=\cE[\varrho_y]=
\cE[\frac{1}{2}\openone]=\frac{1}{2}\openone$ transforms the whole Bloch sphere
into a section of a line containing the total mixture as its center.
Note that $\varrho_x,\varrho_y$ are expressed with respect to the
operator basis with $S_x,S_y,S_z$, i.e. $|\psi\r\l\psi|=P_\psi=\varrho_z$.
In fact, such transformation is completely positive. The matrix $\cE$
takes the form
\be
\cE=\left(
\ba{cccc}
1 & 0 & 0 & 0 \\
\vec{0} & \vec{0} & \vec{0} & \vec{r}^\prime
\ea
\right)\, .
\ee
The singular values of the matrix $E$ are equal to $\lambda_1=\lambda_2=0$
and $\lambda_3=r^\prime$. Consequently, the sufficient criterion for the
complete positivity of unital maps (\ref{unitalsuf}) is fulfilled.

Finally we shall discuss the case when we know that the map under consideration
performs the following transformation of a single input state
\be
\varrho=\frac{1}{2}(\openone+r S_z) \to
\varrho^\prime=\frac{1}{2}(\openone+r^\prime S^\prime_z)\, .
\ee
In this case we must investigate whether the map $\cE$
can be still unital, or whether the total mixture is shifted
from the center. One can easily see that
\be
\nonumber
D(\varrho,\frac{1}{2}\openone)=r\, ; \ \ \
D(\varrho^\prime,\frac{1}{2}\openone)=r^\prime\, ,
\ee
so the unitality depends on the difference $\Delta=r-r^\prime$.
If $\Delta \ge 0$ ($\varrho$ is shifted closer to the center),
then the transformation might still be unital,
but for $\Delta<0$ the quantum channel is inevitable nonunital.

The first guess can be, that we should move the total mixture
as little as possible
so that $D(\cE[\varrho],\cE[\frac{1}{2}\openone])=r$, i.e.
\be
\label{4.7}
\frac{1}{2}\openone\to \cE[\frac{1}{2}\openone]=
\frac{1}{2}\left(\openone+(r^\prime-r)S_z^\prime\right)\, .
\ee
In this case $D(\varrho^\prime,\cE[\frac{1}{2}\openone])
=|(r^\prime-r)-r^\prime|=r$.
Unfortunately, the assumption (\ref{4.7}) about $\cE[\frac{1}{2}\openone]$
is not compatible with the positivity condition.
Note that a linear (not convex) combination
$A=\lambda\varrho+(1-\lambda)\frac{1}{2}\openone=
\frac{1}{2}(\openone+\lambda r S_z)$ for $\lambda=1/r$
corresponds to a pure state $\varrho_z=\frac{1}{2}(\openone+S_z)$.
Applying the knowledge about the state transformation $\cE$
we obtain
\be
\nonumber
\cE[\varrho_z]=\frac{1}{r} \varrho^\prime+\left(1-\frac{1}{r}\right)
\cE[\frac{1}{2}\openone]=
\frac{1}{2}\left(\openone+(1-\Delta)S_z^\prime\right)\, ,
\ee
with $\Delta=r-r^\prime<0$, i.e. the operator $\cE[\varrho_z]$
is not positive, an therefore the map $\cE[\frac{1}{2}\openone]$
cannot be defined in this way. So the shift must be such that
also the operator
$A=\frac{1}{r}\cE[\varrho]+(1-\frac{1}{r})\cE[\frac{1}{2}\openone]$
is positive, i.e. $A=\frac{1}{2}(\openone+\kappa S_z^\prime)$
with $\kappa\le 1$. Comparing two expressions for $A$
we obtain that
\be
\nonumber
\cE[\frac{1}{2}\openone]=
\frac{1}{2}\left(\openone+\frac{r^\prime-\kappa r}{1-r}
S_z^\prime\right)\, .
\ee
The distance
$D(\cE[\frac{1}{2}\openone],\frac{1}{2}\openone)=|(r^\prime-\kappa r)/(1-r)|$
is minimal for $\kappa=1$ (under the condition that $\kappa\le 1$).
Consequently
the guessed state transformation is
\be
\nonumber
\varrho &\to& \varrho^\prime \, ;\\
\varrho_x,\varrho_y,\frac{1}{2}\openone & \to & \cE[\frac{1}{2}\openone]=
\frac{1}{2}\left(\openone+\frac{r^\prime-r}{1-r}S_z^\prime\right)\, .
\nonumber
\ee
We obtain that the pure state $\varrho_z$ is mapped always into
a pure state $\varrho_z^\prime$. It is easy to check that the map
\be
\label{rec1a}
\cE=\left(
\ba{cccc}
1 & 0 & 0 & 0 \\
0 & 0 & 0 & 0 \\
0 & 0 & 0 & 0 \\
\frac{r^\prime-r}{1-r} & 0 & 0 & 1-\frac{r^\prime-r}{1-r}
\ea
\right)
\ee
is indeed completely positive. Note that this is in fact the expression of
the matrix $\cD$ associated with the map $\cE$, i.e. it is not an expression
of the matrix in a fixed basis, but rather a matrix that relates the
two bases $\{S_x,S_y,S_z\}$ and $\{S^\prime_x,S^\prime_y,S_z^\prime\}$.
Our data contain information only about the operators $S_z$ and $S_z^\prime$.
The operators $S_x,S_y,S_x^\prime, S_y^\prime$ are arbitrary, such that
$\T S_j S_k=\T S_j^\prime S_k^\prime \sim \delta_{jk}$, i.e.
in the Bloch-sphere picture they define three mutually orthogonal axes
(a cartesian coordinate system).

The case when $r^\prime\le r$ ($\Delta\ge 0$) is easier.
Then we can make a following guess
\be
\nonumber
\varrho & \to & \varrho^\prime \\
\varrho_x,\varrho_y,\frac{1}{2}\openone & \to &
\cE[\frac{1}{2}\openone]=\frac{1}{2}\openone
\nonumber
\ee
and consequently the map
\be
\label{rec1b}
\cE=\left(
\ba{cccc}
1 & 0 & 0 & 0 \\
0 & 0 & 0 & 0 \\
0 & 0 & 0 & 0 \\
0 & 0 & 0 & r^\prime/r
\ea
\right)
\ee
is completely positive. Note that the matrix
in Eq.(\ref{rec1a}) for $r^\prime=r$ takes the
form (\ref{rec1b}).

In general,  when we do not know how the total mixture is mapped,
our guess can be written in a compact form
\be
\nonumber
\cE=\left(
\ba{cccc}
1 & 0 & 0 & 0 \\
0 & 0 & 0 & 0 \\
0 & 0 & 0 & 0 \\
\max\{\frac{r^\prime-r}{1-r},0\}  & 0 & 0 &
\min\{\frac{r^\prime}{r},1\}-\max\{\frac{r^\prime-r}{1-r},0 \}
\ea
\right)\, .
\ee

\leftline{\bf Note: Adaptable basis   }
Our reconstruction results with the expression of the quantum operation
that is not written in some fixed basis. The matrix $\cE$
acts on the input state written in one basis $\{S_z,S_y,S_z\}$,
but it produces output state written in the basis
$\{S_x^\prime,S_y^\prime,S_z^\prime\}$. Our data determine
only operators $S_z,S_z^\prime$. The choice of other basis operators
must be made so that they are mutually orthogonal. However,
the choice of $S_x^\prime,S_y^\prime$ is irrelevant, because
our map transforms the whole Bloch sphere only into the subspace spanned by
the identity $\openone$ and the operator $S_z^\prime$, i.e.
into the one-dimensional space (line) in the Bloch sphere picture.
The transformation $\cE$ can be expressed in the fixed basis
just by applying a suitable rotation $R_U$, i.e. the unitary transformation
either before, or after the map $\cE$ is applied.

\subsubsection{Examples}

\leftline{\it 1. Identity.}
Let us assume that we find from an experiment that the single input state $\varrho$ is not affected by the
channel. That is, the channel acts on the state $\varrho$ as
an identity $\cI$ and
$\varrho\to\varrho$, i.e. $r^\prime=r$. In this case our  guess of the channel is
\be
\label{exam1a}
\cE=\left(
\ba{cccc}
1 & 0 & 0 & 0 \\
0 & 0 & 0 & 0 \\
0 & 0 & 0 & 0 \\
0  & 0 & 0 & 1
\ea
\right)
\ee
and under the action of this map the Bloch sphere transforms into a line.

\leftline{\it 2. Unitary transformation.}
The case of unitary transformations is basicly the same as the
case of identity, i.e. our guess is represented by the same matrix.
However, we have also information about the transformation of one
operator basis element, namely
\be
\nonumber
\varrho=\frac{1}{2}(\openone+S_z)\to\varrho^\prime=U\varrho U^\prime=
\frac{1}{2}(\openone+S_z^\prime)\, ,
\ee
where $S_x^\prime= \vec{r}^{\,\prime}\cdot\vec{S}$ with $|\vec{r}^{\,\prime}|=1$.
From here it follows that the quantum map $\cE$ can be expressed in the
fixed basis $\{S_x,S_y,S_z\}$ as
\be
\cE=\left(
\ba{cccc}
1 & 0 & 0 & 0 \\
0 & 0 & 0 & x^\prime \\
0 & 0 & 0 & y^\prime \\
0  & 0 & 0 & z^\prime
\ea
\right)\, ,
\ee
where we used a notation $\vec{r}^{\,\prime}=(x^\prime,y^\prime,z^\prime)$.

\leftline{\it 3. Contraction to a pure state.}
Let us assume that from an experiment we know that a test state is transformed by the action of the channel into
the state $P_\psi$:
\be
\varrho\to P_\psi\, .
\ee
Let us consider that $\varrho$ is not pure. Then our guess always
concludes that the map is non-unital.
Moreover, the matrix $\cE$ reads
\be
\label{exam1b}
\cE=\left(
\ba{cccc}
1 & 0 & 0 & 0 \\
0 & 0 & 0 & 0 \\
0 & 0 & 0 & 0 \\
1  & 0 & 0 & 0
\ea
\right)\, ,
\ee
so that all input states are mapped into the state $P_\psi$. It means
that such contraction is perfectly specified (guessed) just based on a single
state transformation. Moreover, we have no doubts that the map $\cE$
has to be a contraction.

The situation is different,
if $\varrho$ is a pure state. Then the estimation (guess) of the map
is expressed by Eq.~(\ref{exam1a}).
That is, the Bloch sphere is mapped into a line
instead into the single point.

\begin{figure}
\centerline{
\includegraphics[width=4cm]{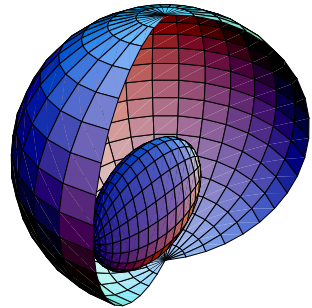}}
\bigskip
\caption{The transformation of the Bloch sphere under the action of the map $\cE$
given by Eq.(\ref{specex}).
The grid on the Bloch sphere and the ellipsoid helps us to see
 the rotation of the original state space. We also see that
the map is non-unital.}
\label{est4}
\end{figure}

\leftline{\it 4. Specific example.}
In this example we will apply our strategy to
a particular channel which in the $\sigma$-basis reads
\be
\label{specex}
\cE=\left(
\ba{cccc}
1 & 0 & 0 & 0 \\
0.5 & 0.2 & -0.1 & 0.1 \\
0 & 0.2 & 0 & -0.3 \\
0 & 0 & 0.3 & 0.3
\ea
\right)\, .
\ee
The action of this map is depicted in Fig.\ref{est4}.

Let us assume that from the experiment we know how a single state
$\varrho=\frac{1}{2}(\openone+0.6\sigma_x)$ is transformed under the action of the
transformation (\ref{specex}), i.e.
\be
\nonumber
\varrho=\frac{1}{2}(\openone+0.6\sigma_x)
\to\varrho^\prime=\frac{1}{2}(\openone+0.62\sigma_x+0.12\sigma_y)\, .
\ee
\begin{figure}
\centerline{
\includegraphics[width=4cm]{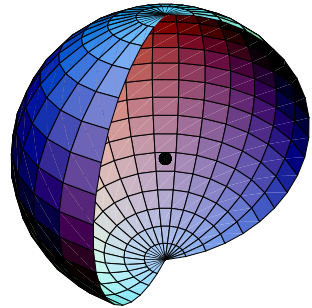}}
\bigskip
\caption{The transformation of the Bloch sphere under the action of the reconstructed
map $\cE_0$, i.e. the contraction of the sphere into a single point representing the total mixture.
Compare with the image of the real map $\cE$ in Fig.~\ref{est4}.}
\label{est0}
\end{figure}
It is easy to verify
that given this transformation
the total mixture has to be shifted,
i.e. the (unknown) channel is non-unital. Simply the inequality
\be
r^\prime=\sqrt{0.62^2+0.12^2}>\sqrt{0.6^2}=r
\ee
implies the non-unitality of the transformation. The reason is that the state
$\varrho^\prime$ lies further from the total mixture than the state
$\varrho$.

In the adaptable basis (see the discussion above) our guess has the form (\ref{rec1a})
\be
\cE^{ad}=\left(
\ba{cccc}
1 & 0 & 0 & 0 \\
0.079 & 0.921 & 0 & 0 \\
0 & 0 & 0 & 0 \\
0 & 0 & 0 & 0
\ea
\right)\, .
\ee
This matrix transforms between two bases: $\{S_j=\sigma_j\}$
and $\{S_x^\prime=\frac{1}{r^\prime}(0.62 \sigma_x+0.12\sigma_y),
S_y^\prime=\frac{1}{r^\prime}(0.12\sigma_x-0.62\sigma_y),
S_z^\prime=\sigma_z\}$, i.e. the output state $\cE^{ad}[\varrho]$
is expressed with respect to the prime basis
$\{S_x^\prime,S_y^\prime,S_z^\prime\}$.
In the $\sigma$-basis the reconstruction
reads
\be
\label{3.24}
\cE_1=\left(
\ba{cccc}
1 & 0 & 0 & 0 \\
0.0776 & 0.904 & 0 & 0 \\
0.015 & 0.175 & 0 & 0 \\
0 & 0 & 0 & 0
\ea
\right)\, .
\ee

\begin{figure}
\centerline{
\includegraphics[width=4cm]{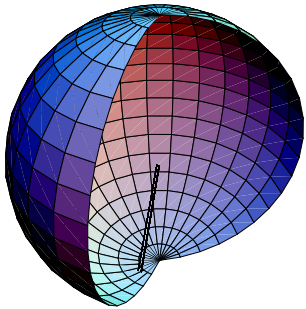}}
\bigskip
\caption{The transformation of the Bloch sphere under the action of the reconstructed
map $\cE_1$ given by Eq.~(\ref{3.24}).
The sphere is transformed into a line. Compare with the image of the real map
in Fig.~\ref{est4}.}
\label{est1}
\end{figure}

\subsection{Two input states}
Let us assume that from an experiment
we know how two input states are transformed under the action of an
unknown map $\cE$. What would be our most reliable reconstruction of the map?

By knowing how two linearly independent input states
are transformed by an unknown map $\cE$ we fix
6 parameters of this map. The  other 6 parameters
have to be specified (guessed). We will  follow our strategy described in the previous section and
we will optimize our guess over six unknown parameters.

\begin{figure}
\label{fig4}
\centerline{
\includegraphics[width=8cm]{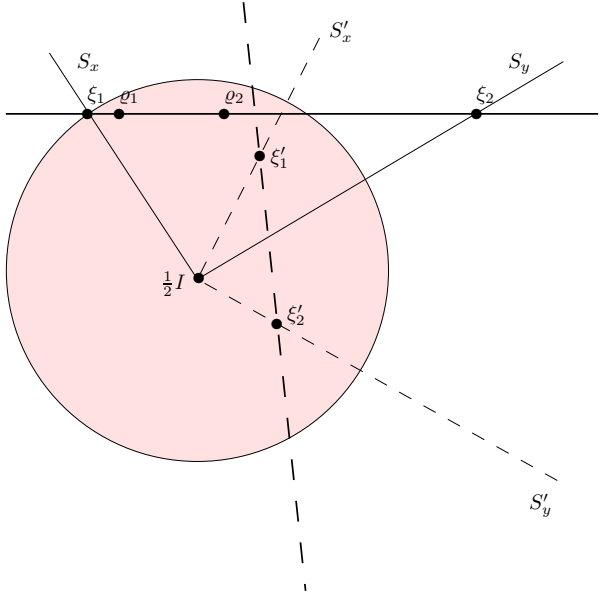}}
\bigskip
\caption{This figure  describes a specific situation of  two known states
$\varrho_{1,2}$.  Solid lines represent the
input  ``cut'' of the Bloch sphere and dashed lines correspond to
the output ``cut'' of the Bloch sphere. Operators
$\xi^\prime_{1,2}$ are images of the states $\xi_{1,2}$, respectively.
Points outside the region of the Bloch sphere
are associated with negative operators with unit trace.
The transformation $\xi_{j}\to\xi_j^\prime$
represents our knowledge about the transformation $\cE$,
i.e. we know that a solid line (given by $\xi_j$)
is mapped into a dashed line (given by $\xi_j^\prime$).
In our reconstruction scheme the initial states
are expressed in $S_j$ basis, whereas the output
states in the $S_j^\prime$ basis.}
\end{figure}

The state transformation that are supposed to be known are:
\be
\label{4.25}
\varrho_1\to\varrho_1^\prime\, , \ \ \ \ \ {\rm and}\ \ \ \ \
\varrho_2\to\varrho_2^\prime\, .
\ee
Thus, we have the information how a two-dimensional subspace
of $\cS(\cH)$ is transformed, or equivalently the
one-dimensional subspace (line) in the Bloch-sphere representation. We can define
two different operators (linear combinations of $\varrho_1$ and $\varrho_2$)
with more suitable properties for the reconstruction.
These new operators are not necessarily positive but they have a  unit trace.
In terms of these new operators we obtain two new transformations
\be
\label{1}
\xi_1=\frac{1}{2}(\openone+S_x) & \to & \xi_1^\prime=
\frac{1}{2}(\openone+\alpha S_x^\prime)\, ; \\
\label{2}
\xi_2=\frac{1}{2}(\openone+b S_y) & \to &
\xi_2^\prime=\frac{1}{2}(\openone+\beta\sin\theta S^\prime_x+
\beta\cos\theta S^\prime_y)\, ,
\ee
that contain the same information about the map $\cE$ as the original ones.
Transformation given by Eq.~(\ref{1}) specifies
how a pure state $\xi_1$ is transformed, while Eq.(\ref{2})
indicates how the
``orthogonal'' (see Fig.~\ref{fig4}) operator $\xi_2$ is transformed.

Let us consider firstly, that the line connecting $\varrho_1$ and
$\varrho_2$ contains a total mixture ($\xi_0=\frac{1}{2}\openone$),
i.e. we know how the total mixture is transformed plus we have
a knowledge about the transformation of one pure state $\xi_1$.
This is the same situation that had been considered in the previous subsection
(now we are certain how the total mixture is transformed).
Following our strategy we put
\be
\nonumber
\xi_0 &=&\frac{1}{2}\openone \to \cE[\frac{1}{2}\openone]
=\frac{1}{2}(\openone+\beta\sin\theta S_x^\prime+\beta\cos\theta S_y^\prime)\,;\\
\nonumber
\xi_1 &=&\frac{1}{2}(\openone+S_x)\to \cE[\xi_1]=\frac{1}{2}(\openone+\alpha S_x^\prime)\, ; \\
\nonumber
\xi_2&=&\frac{1}{2}(\openone+S_y) \to \cE[\frac{1}{2}\openone]
=\frac{1}{2}(\openone+\beta\sin\theta S_x^\prime+\beta\cos\theta S_y^\prime)\, ;\\
\nonumber
\xi_3&=&\frac{1}{2}(\openone+S_z) \to \cE[\frac{1}{2}\openone]
=\frac{1}{2}(\openone+\beta\sin\theta S_x^\prime+\beta\cos\theta S_y^\prime)\, ,
\ee
thus the map $\cE$ reads
\be
\label{e2a}
\cE=\left(
\ba{cccc}
1 & 0 & 0 & 0 \\
\beta\sin\theta & \alpha-\beta\sin\theta & 0 & 0 \\
\beta\cos\theta & -\beta\cos\theta & 0 & 0 \\
0 & 0 & 0 & 0
\ea
\right)\, .
\ee
So under the action of this map  the Bloch sphere transforms into a line.
This map is completely positive
only if $\alpha^2+4\beta^2-4\alpha\beta\sin\theta\le 1$.

Here we have to deal with the self-consistency of the available data, i.e. the transformations
given by Eq.~(\ref{4.25}).
Specifically, not all transformation
$\varrho_1\to\varrho_1^\prime$,
$\varrho_2\to\varrho_2^\prime$ can be simultaneously associated with
completely positive maps. The sufficient and necessary conditions
were derived in Ref.~\cite{Uhlman}. If for all positive $t$ the inequality
\be
D(\varrho_1,t\varrho_2)\ge D(\varrho_1^\prime,t\varrho^\prime_2)
\ee
holds, then there exists a completely positive map $\cE$ such that
$\varrho_j^\prime=\cE[\varrho_j]$ ($j=1,2$). Consequently this condition
must be satisfied by our data. One can also find   simpler necessary
conditions that must hold. For instance the contractivity of the distance (if $t=1$)
of the map $\cE$ has to be fulfilled.

In the case under consideration [see Eq.~(\ref{e2a})] the condition
$D(\frac{1}{2}\openone,\varrho_x)\ge
D(\cE[\frac{1}{2}\openone],\cE[\varrho_x])$ implies that
$\alpha^2+\beta^2-2\alpha\beta\sin\theta\le 1$. Obviously, also $\alpha^2\le 1$
and $\beta^2\le 1$ must hold to ensure that $\cE[\frac{1}{2}\openone]$
and $\cE[\varrho_x]$ are density operators.

If from Eqs.~(\ref{4.25}) we cannot infer the information about the transformation of the
total mixture, then the estimated map takes the form [see Eqs.(\ref{1}-\ref{2})]
\be
\cE=\left(
\ba{cccc}
1 & 0 & 0 & 0 \\
x & \alpha-x & (\beta\sin\theta -x)/b & m \\
y & -y & (\beta\cos\theta-y)/b & n \\
z & -z & -z/b & k
\ea
\right)
\ee
with six free parameters $x,y,z,m,n,k$.
Our strategy is to {\it minimize} the shift of the total mixture (i.e.
$\sqrt{x^2+y^2+y^2}$) and then minimize the distance
$D(\cE[\frac{1}{2}\openone],\cE[\varrho_z])=\sqrt{m^2+n^2+k^2}$
under the condition that the matrix $\cE$ represents a completely
positive map. In principle this task can be performed numerically,
once we have particular values of $b,\alpha,\beta,\theta$. Note that
$\varrho_z=\frac{1}{2}(\openone+S_z^\prime)$, where
$S_z^\prime$ are determined uniquely once $S_x^\prime,S_y^\prime$
are given. Our data contain information about $S_x,S_y,S_x^\prime,S_y^\prime$.
The existence of $\cE$ is not guaranteed for all values of
$\alpha,\beta,\theta$.

Let us first study the case, when it is possible to
set ${\cal E}[\frac{1}{2}{\openone}]={\openone}$,
e.g. the complete mixture is not affected by the map.
In this  case the transformation $\cE$ reads
\[
\cE=\left(
\begin{array}{cccc}
1 & 0 & 0 & 0 \\
0 & \alpha & \beta\sin\theta/b & m \\
0 & 0 & \beta\cos\theta /b & n \\
0 & 0 & 0 & k
\end{array}
\right)\, .
\]
Now we have to determine the range of the parameters involved in this expression for which
the transformation $\cE$ is
completely positive. The contractivity condition of $\cE$
implies that our data must satisfy the relations
\be
\alpha\le 1 \, , \ \ \ \ \ \ \ \ \ \
\beta/b\le 1\, .
\ee
The first inequality has to be always fulfilled whereas the second
one has to be satisfied only if the transformation is unital, i.e. when
$\cE[\frac{1}{2}\openone]=\frac{1}{2}\openone$.
Our guess must satisfy the condition
$\sqrt{m^2+n^2+k^2}\le 1$ because $\cE[\varrho_z]$ must be a density
operator.

In order to proceed further we make an assumption
that $m=n=0$. By applying the transformation $\cE\otimes\cI$
on the maximally entangled state $P_+$ we have to obtain a positive
operator. This guarantees the complete positivity of $\cE$.
In our case, the two smallest eigenvalues
of the system of two qubits are
\[
\mu_{\pm}=\frac{1}{4}\left( 1\pm k-\sqrt{\alpha ^{2}+\left( \frac{\beta }{b}%
\right) ^{2}\pm 2\alpha \frac{\beta }{b}\cos (\theta )}\right) .
\]
From here we derive two conditions
\begin{eqnarray}
1+k &\geq &\sqrt{\alpha ^{2}+\left( \frac{\beta }{b}\right) ^{2}+2\alpha
\frac{\beta }{b}\cos (\theta )}  \label{Condition1} \, ;\\
1-k &\geq &\sqrt{\alpha ^{2}+\left( \frac{\beta }{b}\right) ^{2}-2\alpha
\frac{\beta }{b}\cos (\theta )}\, .  \label{Condition2}
\end{eqnarray}
If we sum both left and right hand sides of Eqs.~(\ref{Condition1}) and (\ref{Condition2}),
we eliminate
the unknown variable $k$ and we obtain the  condition on the input
parameters $\alpha ,b,\beta $ and $\theta $, namely
\begin{equation}
\left( \frac{\beta }{b}\right) ^{2}\leq \frac{1-\alpha ^{2}}{1-\alpha
^{2}\cos (\theta )^{2}}.  \label{Condition3}
\end{equation}
The right hand side of Eq.(\ref{Condition3})
is always smaller than or equal to unity
(equality holds only when the
orthogonality of the input states is preserved). This is in accordance with the
requirement $\beta\le b$. Therefore we can conclude that the inequality (\ref{Condition3})
represents a stronger condition (than $\beta\le b$) that has to be satisfied in order to
preserve the unitality.

Because the (pure) state $\varrho_z$ is transformed into the
state $\varrho_z^\prime=\frac{1}{2}(\openone+k S_z^\prime)$, the
value of $k$ has to be smaller than unity, i.e. $k\le 1$.
Under the given condition (\ref{Condition3}) the value of $k$
can be chosen to be
\be
k=\alpha\sqrt{\frac{1-\alpha^2}{1-\alpha^2\cos^2\theta}}\cos\theta\le 1\, ,
\ee
This value of $k$ fulfills both equations (\ref{Condition1}) and (\ref{Condition2}).
Consequently, under the condition (\ref{Condition3}) the map
$\cE$ is unital and completely positive.
Now our task is to find the minimal value of $k$.
If we try to vary also the parameters $m,n$, then in all numerical tests
the distance $D(\cE[\frac{1}{2}\openone],\cE[\xi_3])$ has been found to be larger than the one
presented above.
Consequently, it is preferable to ``shift'' the third state $\xi_3$
into a state $\xi_3^\prime$ which lies on a line perpendicular
to the plane given by $\frac{1}{2}\openone,\xi_1^\prime,\xi_2^\prime$.

Let us now consider the situation when the total mixture
has to be shifted from the center of the Bloch sphere, i.e.
our guess is such that $\cE[\frac{1}{2}\openone]\ne \frac{1}{2}\openone$.
As we have already mentioned above, in this case we have to optimize
over six free parameters. In the previous subsection (channel reconstruction based
on a measurement of a single test state) the total mixture
has been moved along the line specified by the center of the Bloch sphere
and the given state $\varrho^\prime$. The direct generalization of this
feature  leads us to the following observation: The state
$\cE[\frac{1}{2}\openone]$ lies in the plane determined by points
$\frac{1}{2}\openone,\xi_1^\prime,\xi_2^\prime$.
In what follows we will use the property described at the end of the previous
paragraph: The third state is mapped into a state
that belongs to the line perpendicular to the mentioned plane.
In other words we set $m=n=z=0$ and our guess
takes the form
\be
\cE=\left(
\ba{cccc}
1 & 0 & 0 & 0 \\
x & \alpha-x & (\beta\sin\theta -x)/b & 0 \\
y & -y & (\beta\cos\theta-y)/b & 0 \\
0 & 0 & 0 & k
\ea
\right)\, .
\ee
This apporach reduces our task to a three-parametric problem which we can solve numerically.
We have performed this task and after finding the solution
we were searching in the whole six-parametric space
for transformations that might be better estimates of the map $\cE$. However no such
transformations have been found and therefore we conjecture that
our estimation is the optimal one. Unfortunately, we are not able express explicitly
our guess for general values of $\alpha,\beta,b,\theta$.

\subsubsection{Examples}
\leftline{\it 1. Identity.} Let us assume that the information
available about the action of the map is
of the form
$\varrho_z\to\varrho_z=\frac{1}{2}(\openone+S_x)$
and $\xi_2=\frac{1}{2}(\openone+b S_y)\to
\frac{1}{2}(\openone+bS_y)$. In this case  our guess has the form
\be
\cE=\left(
\ba{cccc}
1 & 0 & 0 & 0 \\
0 & 1 & 0 & 0 \\
0 & 0 & 1 & 0 \\
0 & 0 & 0 & k \\
\ea
\right)\, .
\ee
It is easy to check that the only possibility
is to set $k=1$. Consequently, our reconstruction is perfect
and the identity map is uniquely specified.

The situation is different if we know
how the total mixture is transformed, i.e.
$\cE[\frac{1}{2}\openone]=\frac{1}{2}\openone$.
Then our reconstruction results in
\be
\label{rec2aa}
\cE=\left(
\ba{cccc}
1 & 0 & 0 & 0 \\
0 & 1 & 0 & 0 \\
0 & 0 & 0 & 0 \\
0 & 0 & 0 & 0 \\
\ea
\right)\, .
\ee
This transformation represents
a contraction into the line.

\leftline{\it 2. Unitary transformation.}
The situation is the same as before. Essentially, we have
an information about the transformations
$\varrho_x\to\varrho_x^\prime=\frac{1}{2}(\openone+S_x^\prime)$
and $\xi_2=\frac{1}{2}(\openone+b S_y)\to
\frac{1}{2}(\openone+bS^\prime_y)$. Guessing that
$\frac{1}{2}\openone\to\frac{1}{2}\openone$ we obtain
$S_x\to S_x^\prime,S_y\to S_y^\prime$. Having $S_x,S_y$
then the operator $S_z$ is uniquely identified
(similarly $S_z^\prime$). The reconstructed map $\cE$
has the same form as before except it is not defined in
a fixed basis. This means that our guess is perfect for
unitary transformation. Nevertheless, we have to stress, that the map
under consideration need not be unitary - it can be a non-unital map

The discussion of the case, when we know that
$\frac{1}{2}\openone\to\frac{1}{2}\openone$
and $\varrho_x\to\varrho_x^\prime$ is similar to the case of identity.
The reconstructed map $\cE$ has the form (\ref{rec2aa}) and again it
is not expressed in a fixed operator basis.

\leftline{\it 3. Contraction to a pure state.}

One can efficiently  reconstruct quantum channels of this type. The reason is that
only when $\varrho_x=\frac{1}{2}(\openone+S_x)\to |\psi\r\l\psi|$
one is not able to identify such  channel perfectly with just a single test state.
However, if we have knowledge about two
transformations, our reconstruction of this channel is complete.

\leftline{\it 4. Specific example.}
Let us consider now that we know how  two states
\be
\nonumber
\varrho_1&=&\frac{1}{2}(\openone+0.6\sigma_x)\, ;\\
\nonumber
\varrho_2&=&\frac{1}{2}(\openone+0.4\sigma_x+0.1\sigma_y+0.8\sigma_z)\, ,
\ee
are transformed under the action of the map (\ref{specex})
According to our strategy we firstly construct two perpendicular
(potentially negative) operators  $\xi_1,\xi_2$ as linear
combinations of the states $\varrho_1,\varrho_2$. One of these operators
(e.g., $\xi_1$) is a pure state. We have two options and one of them gives us
the pair
\be
\nonumber
\xi_1&=&\frac{1}{2}(\openone+0.76095\sigma_x-0.080475\sigma_y-0.6438\sigma_z)\, ;\\
\xi_2&=&\frac{1}{2}(\openone+0.464776\sigma_x+0.0676122\sigma_y+0.540897\sigma_z)\, .
\nonumber
\ee
We see that the operator $\xi_2$ is  negative, i.e. it does not
correspond to any quantum state and lies outside
the Bloch sphere. The output states
can be expressed as
\be
\nonumber
\xi_1^\prime&=&\frac{1}{2}(\openone+0.595857\sigma_x+0.34533\sigma_y-0.217283\sigma_z)\, ;\\
\xi_2^\prime&=&\frac{1}{2}(\openone+0.6402284\sigma_x-0.0693141\sigma_y+0.182553\sigma_z)\, .
\nonumber
\ee
and one can easily evaluate the required parameters
\be
\nonumber
b &=& ||\vec{\xi}_2||=0.71635 \, ;\\
\nonumber
\alpha & = & ||\vec{\xi^\prime}_1|| = 0.722157\, ; \\
\nonumber
\beta &=& ||\vec{\xi^\prime}_2|| = 0.669398\, ; \\
\nonumber
\theta &=& \arcsin \left(\frac{1}{\alpha\beta}\vec{\xi_1^\prime}\cdot\vec{\xi^\prime}_2\right)=0.717699 {\rm\ rad} \, .
\ee
where we have used the notation $\xi_j=
\frac{1}{2}(\openone+\vec{\xi}_j\cdot\vec{\sigma})$. The above formulae
follow from the expressions for $\xi_1,\xi_2\xi_1^\prime,\xi_2^\prime$
in Eqs. (\ref{1}) and (\ref{2}), i.e.
\be
\nonumber
\xi_1&=&\frac{1}{2}(\openone+S_x)\, ; \\
\nonumber
\xi_2&=&\frac{1}{2}(\openone+b S_y)\, ; \\
\nonumber
\xi_1^\prime &=& \frac{1}{2}(\openone+\alpha S_x^\prime)\, ; \\
\xi_2^\prime &=& \frac{1}{2}(\openone+\beta\sin\theta S_x^\prime+\beta\cos\theta S_y^\prime)\, .
\nonumber
\ee
These relations help us to construct the rotation matrices $X,Y$ that transform
the basis $\sigma_j\leftrightarrow S_j$ into
$\sigma_j\leftrightarrow S_j^\prime$, and vice verse. Our reconstruction results
in the matrix written in the adaptable basis, which transforms
the matrices written in the $S$-basis into the $S^\prime$-basis. The expression of this map
in the $\sigma$-basis is obtained via the relation $\cE_2= Y^{-1} \cE^{ad} X$

\begin{figure}
\centerline{
\includegraphics[width=4cm]{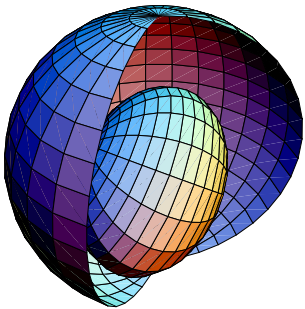}}
\bigskip
\caption{This figure illustrates the transformation of the original
Bloch sphere under the action of the reconstructed
map $\cE_2$ given by Eq.~(\ref{3.52}.
The Bloch is transformed into an ellipsoid. We can   compare this
ellipsoid  with the ellipsoid obtained under the action of the original map presented in Fig.(\ref{est4}).}
\label{est2}
\end{figure}

From above it follows that
the map is not unital and therefore, we have to search
numerically for the solution. As a result we find
\be
\nonumber
\cE^{ad}=\left(
\ba{cccc}
1 & 0 & 0 & 0 \\
0.101568 & 0.620589 & 0.472761 & 0 \\
0.0600669 & -0.0600669 & 0.620094 & 0 \\
1 & 0 & 0 & 0.457
\ea
\right)\, .
\ee
Using the two unitary transformations $X,Y$ we obtain the guess (the reconstruction of the map $\cE$)
in the fixed basis
\be
\label{3.52}
\cE_2=\left(
\ba{cccc}
1 & 0 & 0 & 0 \\
0.1168 & 0.83866 & -0.03137 & 0.25083 \\
0.01523 & 0.1746 & 0.26946 & -0.34023 \\
0.00696 & -0.0116 & 0.36878 & 0.28862
\ea
\right)\, .
\ee

\subsection{Three input states}
Given three input states we know how a whole plane is transformed.
Again we have two possibilities: either this plane contains
the total mixture, or not. If yes, then we have the knowledge about
the following transformations
\be
\nonumber
\xi_0&=&\frac{1}{2}\openone \to \xi_0^\prime\,
=\frac{1}{2}(\openone+x S_x^\prime +y S_y^\prime)\, ;\\
\nonumber
\xi_1&=&\frac{1}{2}(\openone+S_x)\to\xi_1^\prime
=\frac{1}{2}(\openone+\alpha S_x^\prime )\, ;\\
\nonumber
\xi_2&=&\frac{1}{2}(\openone+S_y) \to \xi_2^\prime
=\frac{1}{2}(\openone+\beta\sin\theta S_x^\prime+\beta\cos\theta S^\prime_y)\, .
\ee
Like before, we assume that our experimental data
are given as $\varrho_j\to\varrho_j^\prime$ ($j=1,2,3$) and
we express the available information in
a form more suitable for our purposes
via the states $\xi_j$.

Our aim is to find $m,n,k$ such that the distance
$D(\xi_0^\prime,\xi_3^\prime)$ is minimal while
\be
\nonumber
\xi_3 =\frac{1}{2}(\openone+S_z) \to \xi_3^\prime
=\frac{1}{2}(\openone+m S_x^\prime+n S^\prime_y +k S_z^\prime)\, .
\ee
Of course, the estimated transformation $\xi_3\to\xi_3^\prime$
has to fulfill the conditions of the complete positivity.
The situation is similar as the one discussed in the previous subsection.
The only difference is that now we are sure how the total mixture
is transformed, i.e. whether it stays in the center of the Bloch sphere,
or not.

In the second case, when the plane does not contain the total mixture,
we are forced to make an estimation about its position in the state space.
It is impossible to solve the problem
in the most general case for all possible parameters. However, in particular cases
(having numerical values of given parameters), we can easily estimate the
transformation $\cE$ in accordance with our strategy.

\subsubsection{Examples}

\leftline{\it 1. Identity.} In all cases the identity will be identified perfectly.
If we do not know how the identity is transformed, then the guess
$\frac{1}{2}\openone\to\frac{1}{2}\openone$ is reasonable. In fact, we have no
other choice. Similarly, if we know that
$\frac{1}{2}\openone\to\frac{1}{2}\openone$, then our only possibility is to
set $\xi_3\to\xi_3^\prime=\xi_3$.

We know that the map is unital, i.e.
$\frac{1}{2}\openone\to\frac{1}{2}\openone$ and for unital maps
with $\lambda_1=\lambda_2=1$ the necessary conditions are very strict
\be
1+k\ge 2\, ,
\ee
and imply that $k=1$.

\leftline{\it 2. Unitary transformation.} In this case
we are dealing essentially with  the same situation as in the case of the identity.
Knowing the transformation of  three states our guess of the unitary transformation
is perfect.

\leftline{\it 3. Contraction to a pure state.} As discussed in the previous section this map can be
completely determined based on the knowledge of how two linearly independent states are transformed.

\leftline{\it 4. Specific example.}
Let us assume that transformations of the following three states
are given (known from the measurement)
\be
\nonumber
\varrho_1&=&\frac{1}{2}(\openone+0.6\sigma_x)\, ;\\
\nonumber
\varrho_2&=&\frac{1}{2}(\openone+0.4\sigma_x+0.1\sigma_y+0.8\sigma_z)\, ;\\
\nonumber
\varrho_3&=&\frac{1}{2}(\openone+0.4\sigma_x+0.3\sigma_y+0.6\sigma_z)\, .
\ee
Now we are left with only three free  parameters and we have to search for
the reconstruction numerically. The result of our numerics is
\be
\label{e3}
\cE_3=
\left(
\ba{cccc}
1 & 0 & 0 & 0 \\
0.294686 & 0.54219 & -0.02396 & 0.176042 \\
0 & 0.2 & 0 & -0.3 \\
0.0562144 & -0.0936907 & 0.27918 & 0.27918
\ea
\right)\, .
\ee

\begin{figure}
\centerline{
\includegraphics[width=4cm]{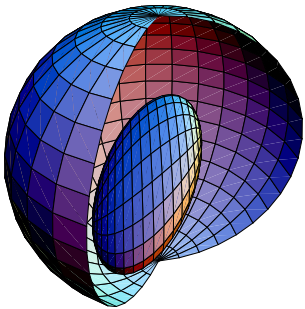}}
\bigskip
\caption{The picture  of the  Bloch sphere that is transformed by the action of the reconstructed
map $\cE_3$ given by Eq.~(\ref{e3}). Compare with the transformation of the Bloch sphere  by the  map $\cE$
in Fig.~\ref{est4}.}
\label{est3}
\end{figure}

\section{Discussion and Conclusions }
In this paper we have presented  strategies how to reconstruct (estimate)
properties of a quantum channel described by the map $\cE$. In a particular case
of a qubit channel a complete reconstruction of the map $\cE$ can be performed
via complete tomography of four output states $\cE[\varrho_j]$ that form at the input
of the channel a set of four linearly independent states $\varrho_j$ ($j=1,2,3,4$).
We have studied the situation when less than four linearly independent states are transmitted
via the channel and measured at the output. We have presented strategies how to reconstruct
the channel when just one, two and three states are transmitted via the channel.
We have shown that unitary transformations (channels) can be
uniquely reconstructed (determined) based on the information of how three properly chosen input states
are changed under the action of the channel.
We conclude the paper with three remarks. Firstly we will comment on the channel capacity associated
with reconstructed (estimated) maps. Secondly, we will address the problem of optimality of the
channel estimation.

\subsection{ The channel capacity.}

Let us study how the channel capacity
defined by the expression (for details see, e.g. Ref.~\cite{Nielsen})
\be
C=\max_{\varrho=\sum_j p_j\varrho_j}[S(\varrho)-\sum_j p_j S(\varrho_j)]
\ee
depends on reconstruction strategies.

Prior any measurement on the channel is performed, the conservative assumption about the
channel capacity should be that it is zero,
i.e. the channel is ``useless'' for the transmission
of classical information. One might expect that any measurement performed on the channel
(i.e. via sending a specifically prepared state through the channel and measuring it at the
output) should result in the estimation of the channel that has capacity closer to the capacity
of the actual channel.
However, in general this is not always the case. To illustrate this let us assume the following problem:
Consider that we are trying to guess
the capacity of the contraction to a pure state $P_\phi$. If our
experiment says that $P_\psi\to P_\phi$, then the guessed
transformation preserves the total mixture, and transforms
the whole state space (the Bloch sphere) into a line instead into the point. Therefore,
the capacity of the first estimation is non-vanishing, whereas
the actual channel has the zero capacity.

\leftline{\it Unital channels}
Let us consider  an estimation of a unital channel. The
the capacity of a unital channel is given by the expression \cite{King}
\be
C(\cE)=1-H(\mu)\, ,
\ee
where $H(\mu)=-\frac{1}{2}(1-\mu)\log\frac{1}{2}(1-\mu)-\frac{1}{2}(1+\mu)\log\frac{1}{2}(1+\mu)$,
$\mu=\max\{|\lambda_1|,|\lambda_2|,|\lambda_3|\}$
and $\lambda_j$s are singular values of the matrix $E$. The value
of $\mu$ is related to the maximal distance between the total
mixture and some state $\varrho^\prime=\cE[\varrho]$. Because in our estimation strategy
we are searching for maps that minimize such distance, we
always reconstruct a channel which has lower (or equal) capacity as the actual
channel. As a result we obtain that for unital maps
our estimation strategy is in accordance with the channel capacity approach: Better our estimation is
closer is the estimated channel capacity to the capacity of the actual channel.

\subsection{Fidelity of channel estimation}
In order to quantify the fidelity of the channel estimation we have to introduce
a corresponding measure. The average fidelity between a map $\cE$ and some reference map $\cE_r$
can be quantified as an integral
\be
\label{integ}
\cF = \int d\cE [d(\cE,\cE_r)]\, ,
\ee
over all possible maps $\cE$. Unfortunately, we do not know how to specify a proper integration measure
on the space of CP maps $\cE$.
For this reason it is much easier to consider an average distance between two CP maps $\cE$ and $\cG$ that is defined
as
\be
\label{distance}
d(\cE,\cG)=\int_{\cS(\cH)}[D(\cE[\varrho],\cG[\varrho])]d\varrho \, .
\ee
where the average is performed over whole state space of the system on which the maps do act.

A good reconstruction scheme has a property that the update of our information cannot
debase our estimation. In particular, it means that $\cE_1$ is better
estimate as $\cE_0$, $\cE_2$ is better than $\cE_1$, etc, where $\cE_n$ represents
our guess with $n$ known state transformations ($n=0,1,2,3,4$).
Using the average distance (\ref{distance}) this property can be formally expressed
via a sequence of inequalities
\be
\label{hierarchy}
d(\cE_0,\cE)\ge
d(\cE_1,\cE)\ge
d(\cE_2,\cE)\ge
d(\cE_3,\cE)\ge
d(\cE_4,\cE)\, ,
\ee
where $\cE$ is the actual completely positive map that is estimated and
$\cE_j$ are corresponding estimates.

In our case $\cE_0$ is a contraction of the whole state space
into the total mixture. We can evaluate explicitly the average distance (\ref{distance}) in this case
\be
\nonumber
d(\cE_0,\cE)=\int d\varrho [\T|\frac{1}{2}\openone-\cE[\varrho]|]=
\int_{\cE[\cS(\cH)]} d\vec{r} |\vec{r}|\, ,
\ee
where we have used the property
$D(\frac{1}{2}\openone,\varrho)=|\vec{r}|$. Consequently,
$d(\cE_0,\cE)$ corresponds to a mean distance of $\cE[\cS(\cH)]$
to the center of the Bloch sphere.

Due to the triangle inequality
\be
\nonumber
D(\cE_0[\varrho],\cE[\varrho])+
D(\cE_0[\varrho],\cE_1[\varrho])\ge D(\cE_1[\varrho],\cE[\varrho])
\ee
we find that
\be
\nonumber
d(\cE_0,\cE)+d(\cE_1,\cE_0)\ge d(\cE_1,\cE)\, .
\ee
The fact that image of the Bloch sphere under the action of the map  $\cE_1$ is a line
(a set of measure zero) implies that
\be
\nonumber
d(\cE_0,\cE_1)=\int_{\cE_1[\cS(\cH)]}d\vec{r}|\vec{r}|=0
\ee
and
\be
d(\cE_0,\cE)+d(\cE_1,\cE_0)=d(\cE_0,\cE)\ge d(\cE_1,\cE)\, .
\ee
From here it follows that
the relation $d(\cE_0,\cE)\ge d(\cE_1,\cE)$
holds. Consequently the estimation $\cE_1$ is better than
the estimation $\cE_0$. This relation is in accordance with our intuition:
A guess based on some data must be always better than
a random guess.

Following the above argument  we conjecture that
the whole hierarchy  of inequalities (\ref{hierarchy}) holds. This would mean that
larger the set of test states better our
estimation  is.

Let us
consider the specific example of the map (\ref{specex}) that has been studied throughout Sec.~III.
This is a non-unital map for which we have presented its estimations in various situations, i.e. in cases
when one, two or three test states have been sent via the channel. Using the corresponding estimates $\cE_j$ we
can evaluate the average distances $d(\cE_j, \cE)$ for which we find
\be
\nonumber
d(\cE_0,\cE)&\approx& 0.845 \, ; \\
\nonumber
d(\cE_1,\cE)&\approx& 0.758 \, ; \\
\nonumber
d(\cE_2,\cE)&\approx& 0.636 \, ; \\
\nonumber
d(\cE_3,\cE)&\approx& 0.348 \, .
\ee
We can conclude that the hierarchy of inequalities (\ref{hierarchy}) for this specific non-unital map is preserved.
Once this hierarchy is proved to be valid, one can ask a question how to chose the set of
test states $\varrho_j$ so that the
sequence of distances (\ref{hierarchy}) converges to zero (the perfect estimation) most rapidly.

\subsection{Optimal test states}
One of our aims was to investigate which states
are  efficient for the process reconstruction.
It turns out that it is reasonable to start with the
total mixture  as the first test state. Starting with a pure state,
the first guess is always unital. Using a general mixed state
the shift of the center of the Bloch sphere is only estimated, but
if $\varrho_1=\frac{1}{2}\openone$, then the question of unitality
is solved without any doubts. Therefore, we suggest to use the
total mixture as the first test state. In this case,
the channel capacity always vanishes after the first step
of estimation.
To improve the reconstruction one has to send via the channel
more states. The best strategy is to complement the total mixture with
pure states which are mutually orthogonal in the sense
of Bloch sphere representation.
This optimization of
the reconstruction via the choice of test states is still an open question.

\begin{acknowledgments}
This work was supported by the European Union  projects
QGATES and CONQUEST.
\end{acknowledgments}


\end{multicols}

\end{document}